\documentclass{cupconf}
\input psfig.sty

  \checkfont{eurm10}
  \iffontfound
    \IfFileExists{upmath.sty}
      {\typeout{^^JFound AMS Euler Roman fonts on the system,
                   using the 'upmath' package.^^J}%
       \usepackage{upmath}}
      {\typeout{^^JFound AMS Euler Roman fonts on the system, but you
                   dont seem to have the}%
       \typeout{'upmath' package installed. cupconf.cls can take advantage
                 of these fonts,^^Jif you use 'upmath' package.^^J}%
      }
  \else
  \fi

  \checkfont{msam10}
  \iffontfound
    \IfFileExists{amssymb.sty}
      {\typeout{^^JFound AMS Symbol fonts on the system, using the
                'amssymb' package.^^J}%
       \usepackage{amssymb}%

      }{}
  \fi

  \IfFileExists{amsbsy.sty}
    {\typeout{^^JFound the 'amsbsy' package on the system, using it.^^J}%
     \usepackage{amsbsy}}
    {}

\newsavebox{\astrutbox}
\sbox{\astrutbox}{\rule[-5pt]{0pt}{20pt}}

\newcommand\arcdeg{\mbox{$^\circ$}}%

\title[Exoplanet Opportunities]{Observations of Extrasolar Planets Enabled by a Return to the Moon}

\author[P. R. McCullough]%
{P\ls E\ls T\ls E\ls R\ns M\ls c\ls C\ls U\ls L\ls L\ls O\ls U\ls G\ls H$^1$}

\affiliation{$^1$Space Telescope Science Institute, 3700 San Martin Dr., Baltimore MD 21218}

\pubyear{2007}
\volume{XXX}
\pagerange{001--008}
\date{?? and in revised form ??}
\begin{document}

\maketitle

\begin{abstract}
Ambitious studies of Earth-like extrasolar planets are outlined in the context
of an exploration initiative for a return to the Earth's Moon. 
Two mechanism for linearly polarizing light reflected from Earth-like planets are discussed: 1)
Rayleigh-scattering from a planet's clear atmosphere, and
2) specular reflection from a planet's 
ocean. Both have physically simple and predictable polarized phase functions.
The exoplanetary diurnal variation of the polarized light reflected from a
ocean but not from a land surface 
has the potential to enable reconstruction of the continental boundaries on
an Earth-like extrasolar planet.
Digressions on the lunar exploration initiative also are presented.
\end{abstract}

\firstsection 
\section{Introduction}
\label{sec:intro}

For millennia humans have observed the classical planets as unresolved
points of light moving on the celestial sphere.  Now we are on the
verge of seeing {\it extrasolar} planets as unresolved, moving points of
light. We also are exploiting, or soon will exploit, alternative methods
of imaging extrasolar planets, one of which we propose in this paper:
sea-surface glints.

Astronomical research and human exploration have a strong historical
precedent.  The global oceanic explorations of Captain James Cook and
others were in large part motivated by the possibility of measuring the
scale of the solar system in physical units, e.g. meters, by geometrical
triangulation of the silhouette of Venus transiting the Sun as viewed from
nearly opposite sides of the Earth during the transit of Venus of 1769
(Maor 2000).  Cook's expedition required the significant resources of the
greatest naval power on Earth to sail for nearly 8 months from England
to Tahiti. That is, he journeyed as far as was humanly possible in the
18th century in order to attempt to answer a scientific question as old
as humankind, ``How far is the Sun, and hence by elementary geometry,
how far are the other planets?''  Although Cook and others were blessed
with cloud-free skies on the day of the transit, the scientific objective
of his expedition failed due to turbulence in the Earth's atmosphere.
From the long view of history, we do not recall such expeditions' costs
in lives or gold, nor does it matter much that they failed in their
scientific objectives.  Instead, we revere such expeditions for their
power to inspire humans to collectively seek great accomplishments for
the good of all humankind.

In this paper, we outline a similar scientific quest to understand
humankind's place in the cosmos: a historically significant and achievable
scientific goal for NASA would be to identify {\it oceanic} extrasolar
planets.  This goal is simpler than, and also related to, the goal of
former NASA administrator Dan Goldin to image the surfaces of extrasolar
planets.  If we were to attempt such imagery by diffraction-limited
optics, the minuscule angular size of an extrasolar planet would require
truly enormous separations between the optical components. However,
an indirect imaging technique such as that presented in Section
\ref{sec:glints} may permit mapping the surfaces of extrasolar planets
with telescopes that are well within the capabilities of current human
technologies. In Section \ref{sec:context} I provide some context to the
lunar exploration initiative and some digressions on possible
impacts it may have on society.

\begin{figure}
\psfig{figure=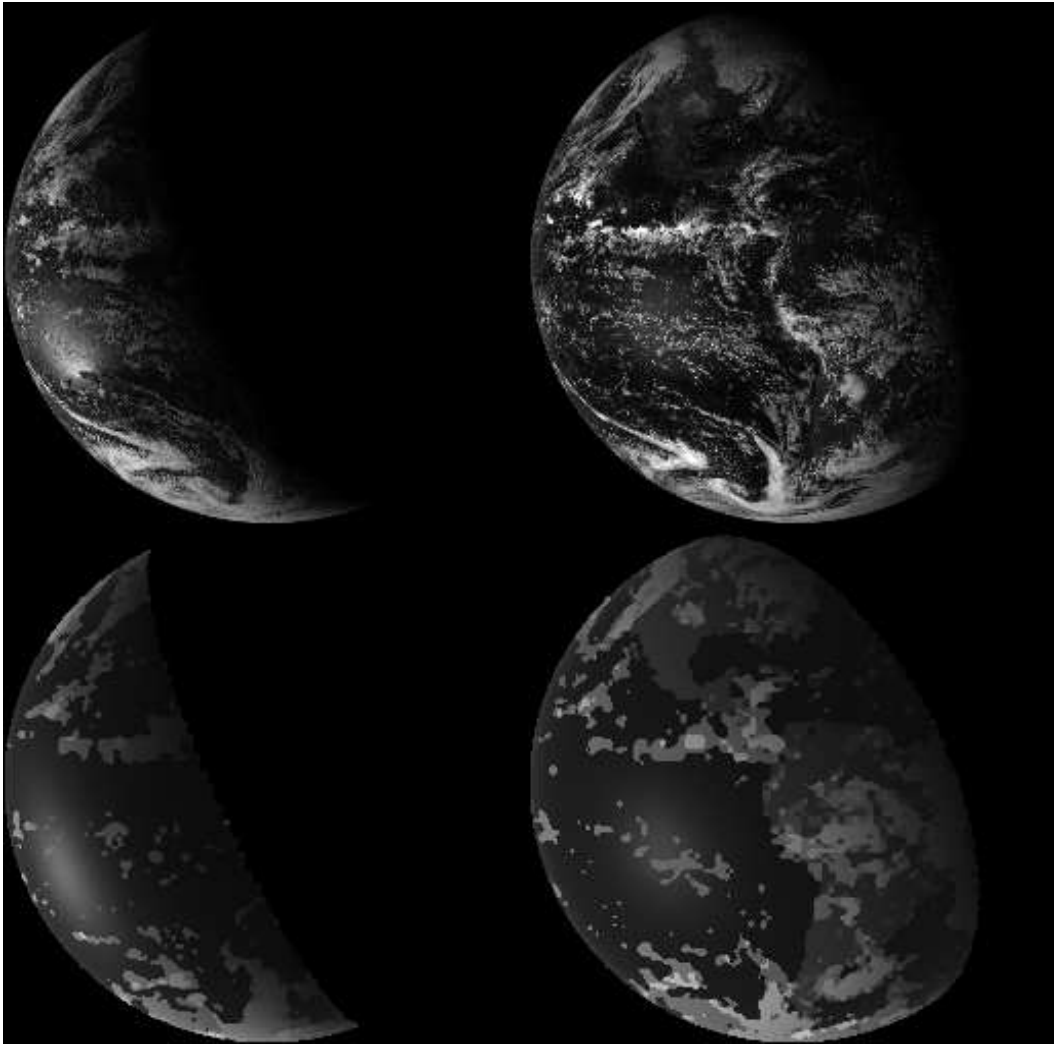,angle=0,width=5.5truein}
\caption{Images from a geostationary satellite (above) taken
in unpolarized light 
and corresponding numerical models for the same (below), from McCullough (2007).
The phase angle $\gamma = 100$\arcdeg\ (left) and 59\arcdeg\ (right).
The sea-surface glint is the bright (white) patch to the lower-left of center of
the Earth.
The Americas are visible in the image and model at right.
The slight brightening along the left edge of the Earth where
there are no clouds is due to Rayleigh scattering in the atmosphere.
\label{fig:goesnone}}
\end{figure}

\begin{figure}
\psfig{figure=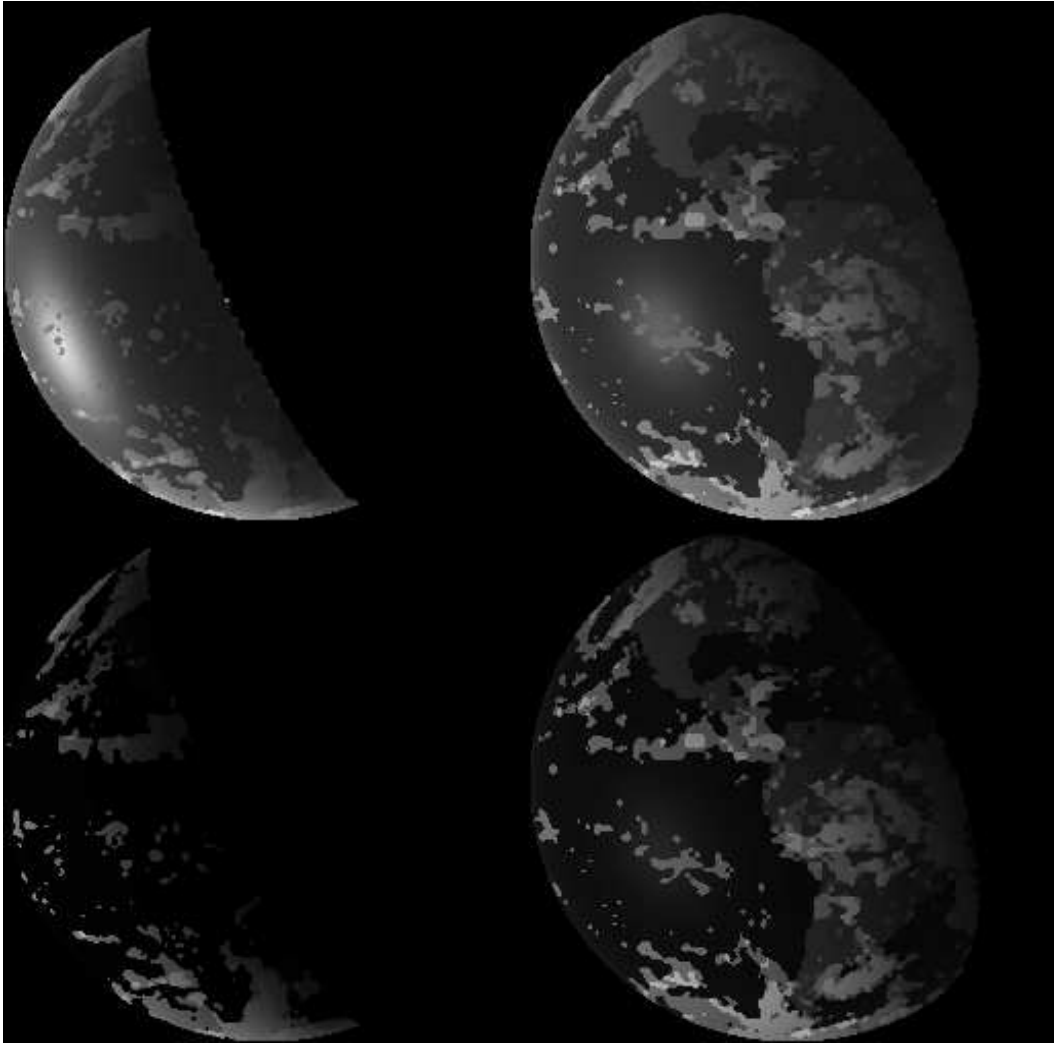,angle=0,width=5.5truein}
\caption{Models images of the scenes from Figure \ref{fig:goesnone}
are shown for linear polarized components of light
($\perp$\ or s-component above; $\parallel$\ or p-component below),
from McCullough (2007).
\label{fig:goespolarized}}
\end{figure}

\begin{figure}
\psfig{figure=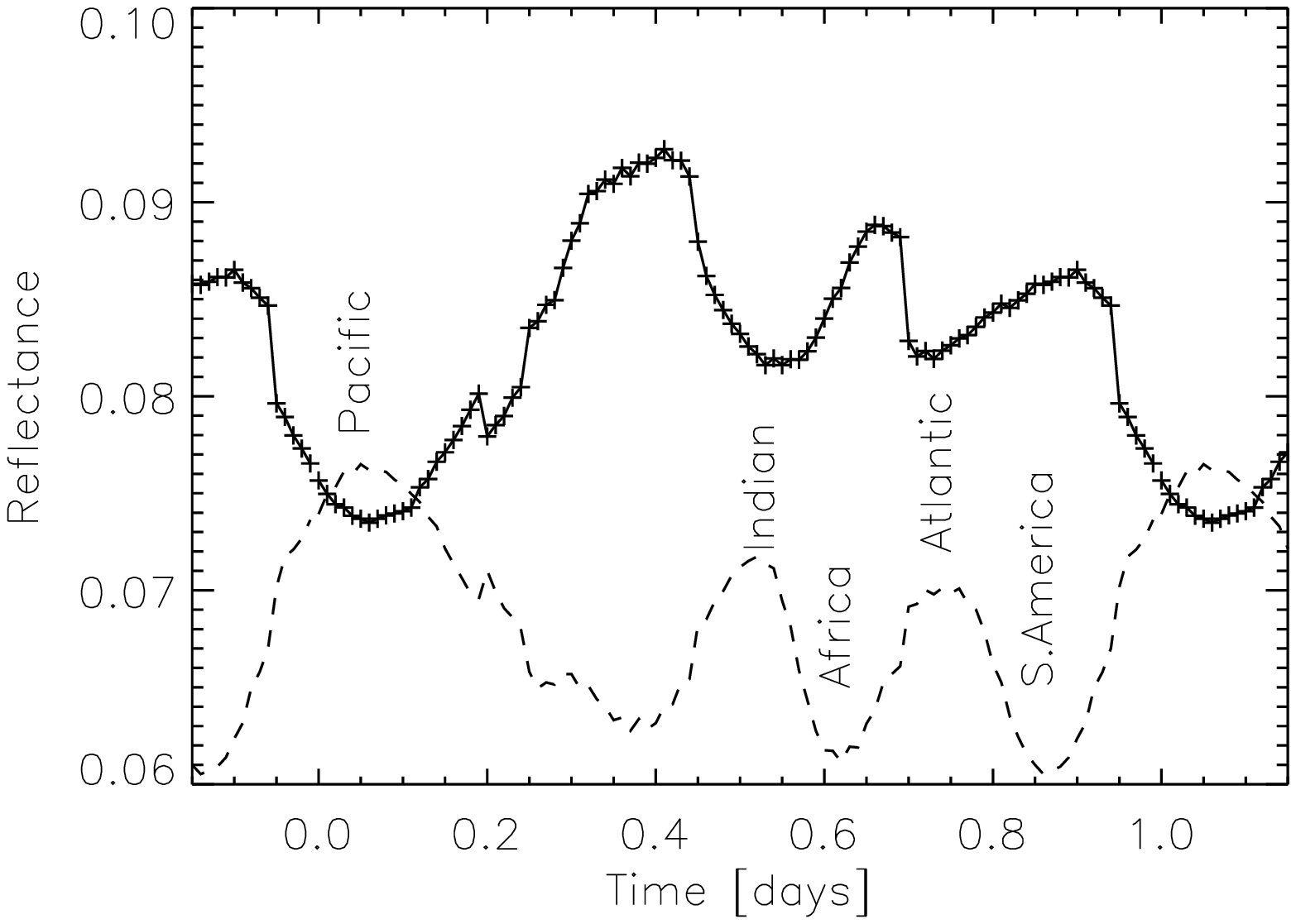,angle=0,width=5.5truein}
\caption{Diurnal light curves for the Earth viewed at quadrature, from
McCullough (2007).
The reflectance ($+$ symbols) includes contributions from land, sea,
atmosphere and authentic clouds;
it is normalized to 1.0 for a full-phase Lambertian sphere.
Twice the difference between the s- and
p-components of linear polarization (dashed line) is 
anti-correlated with
the total reflectance, because clouds increase the total reflectance while
attenuating the two mechanisms for polarization, glint from the ocean
and Rayleigh scattering of the clear atmosphere.
Major surface features at the location where
the sea-surface glint is or would be, in the case of continents, are
labeled above the dashed line: the local maxima correspond to clear skies
over oceans (Pacific, Indian, and Atlantic, as labeled);
the local minima, to continents (Africa and South America, as
labeled) or cloudy regions (between the Pacific and Indian oceans).
\label{fig:fst}}
\end{figure}

\section{Glints from the Oceans of Extrasolar Planets}
\label{sec:glints}

We need not spatially resolve an extrasolar planet in
order to determine if it has an atmosphere and an ocean like those
of Earth.  Instead, we propose to exploit the linear polarization
generated by Rayleigh scattering in the planet's atmosphere and specular
reflection (glint) from its ocean to study Earth-like extrasolar planets.
In principle we can map the extrasolar planet's continental boundaries by
observing the glint from its oceans periodically varying as the rotation
of the planet alternately places continents or water at the location
on the sphere at which light from the star can be reflected specularly
to Earth.

The concepts in this section have been described by McCullough (2007)
and independently by others.  Seager et al. (2000), Saar \& Seager
(2003), Hough \& Lucas (2003), Stam, Hovenier, \& Waters (2004), and
Stam \& Hovenier (2005) have examined the Rayleigh-scattered light
of a hot Jupiter.  Williams \& Gaidos (2004) and Gaidos et al. (2006)
examined the unpolarized variability of the sea-surface glint from an
Earth-like extrasolar planet.  Stam \& Hovenier (2006) independently
examined the observability of the polarized signatures of an Earth-like
extrasolar planet, including Rayleigh scattering and sea-surface glint.
Kuchner (2003) and L{\'e}ger et al. (2004) have proposed that ``ocean
planets'' may form from ice planets that migrate inward and melt; the
surfaces of these planets would be liquid water exclusively, i.e. no
continents, and it remains to be seen whether such planets would be
entirely obscured by a thick steam atmosphere, or instead might have
clear atmospheres like Earth's.

Specularly reflected light, or glint, from an ocean surface may provide a
useful observational tool for studying extrasolar terrestrial planets.
An interesting parallel exists between using glints to image the
oceans of extrasolar planets and a similar technique to image within
the Earth's turbid ocean (e.g. Moore et al. 2000). In the latter, one
aims an underwater camera toward the sea floor while illuminating the
camera's field of view with a laser beam.  Laser light scattered by the
ocean water creates a haze of light visible by the camera, but the laser
light that reaches the sea floor creates a well-defined spot that is also
detected by the camera. By scanning the laser across the sea floor and
simultaneously recording the location and brightness of the peak of the
image, the light scattered by the turbid water is suppressed and detection
of objects on the sea floor is enhanced.  In the proposed technique for
imaging extrasolar planets, the glint acts like the localized spot of the
laser beam, and the rotation of the planet under the glint serves much
the same purpose as the scanning of the laser beam.  The polarization
of the glint allows one to isolate it from the less-polarized reflection
from other regions on the surface, and from the nearly-unpolarized star
light scattered in the telescope optics. In the underwater technique,
the monochromatic color of the laser's light can be used to increase
its contrast over any ambient light.\footnote{Also, the laser can be
pulsed.} Analogously, the two primary mechanisms of polarization, the
glint and atmospheric Rayleigh scattering, can be differentiated with
color, since the former is nearly achromatic whereas the latter depends
strongly on wavelength.

Detection of sea-surface glints would differentiate ocean-bearing
terrestrial planets, i.e. those similar to Earth, from other terrestrial
extrasolar planets.  The brightness and degree of polarization of
both sea-surface glints and atmospheric Rayleigh scattering are strong
functions of the phase angle of the extrasolar planet (see McCullough
2007).  The difference of the two orthogonal linearly polarized
reflectances may be an important observational signature of Rayleigh
scattering or glint.  The difference attributable to Rayleigh scattering
peaks near quadrature, i.e. near maximum elongation of a circular orbit.
The difference attributable to glint peaks in crescent phase, and in
crescent phase the total (unpolarized) reflectance of the glint also
is maximized. The reader intrigued by this short summary will find
additional detail in McCullough (2007) and references therein.

\section{Context and Digressions}
\label{sec:context}

As quoted in his son's book (Dyson 2002), Freeman Dyson wrote in May 1958, 
\begin{quotation}
We shall know what we go to Mars for, only after we get there.... You might
as well ask Columbus why he wasted his time discovering America when he
could have been improving the methods of Spanish sheep-farming. It is lucky
that the U.S. Government like Queen Isabella is willing to pay for the ships.
\end{quotation}
In the case of the Earth's Moon, we have been there already, and
to me, the potential tangible benefits of lunar exploration are not as
clear as the intangible benefit of inspiring people to reach collectively
for grand accomplishments. Preparing for this conference has
caused me to consider many things that I wouldn't have otherwise.
In this section I digress to consider (in order) the relative cost of NASA,
some potential risks of the US not exploring the Moon, a significant
budgetary challenge to sustaining that initiative, some benefits
and concomitant risks of the lunar-exploration initiative.

Humans commonly redefine large quantities in appropriate units, such
as the A.U. for solar system distances, or the parsec for stellar
distances. I need to do the same for the large sums of money associated
with lunar exploration.  Financially, NASA is approximately equivalent
to a single large corporation.  The median market capitalization of
the 30 corporations in the Dow Jones Industrial Average is 108 B USD
(as of Oct 31, 2006). One such corporation, Pfizer, a pharmaceutical
company founded in 1849 and headquartered in NYC, in 2005 had annual
revenue of 51 B USD and spent 7 B USD on R\&D.  NASA by comparison was
awarded 16 B USD in 2005 by the US Congress.

The net worth of US households in 2000 was 50 T USD\footnote{In this and
the other examples, I quote financial figures for the US because they
were readily available to me on the Web (Yardeni 2004) and because this conference is
primarily addressing the US-NASA initiative to return to the Moon.}.
The aggregate value of corporate equities directly held was 9 T USD in
2000 and was 4 T USD in 2003, so it declined approximately 2 T USD per
year for three consecutive years. That is 2000 B USD per year, or 1 B USD
per hour of each and every working day for three years. Although it is
hard for me as an individual to grasp these large sums, these comparisons
may help put into context that NASA's lunar exploration initiative is
expensive in one sense but not relative to the richness and power of our
society, and especially not in comparison to those of the global society.

Returning to the Moon may make sense
when one considers the potential risks or costs to the
USA of {\it not} returning to the Moon when other nations do.
Each nation may have a self-interest in establishing a presence on the Moon,
much like Antarctica,
in order to assure that no single nation monopolizes it.
As someone\footnote{Regretably, I cannot recall the exact quotation or who said it.} once said when asked how much the US should spend on science, 
\begin{quotation}
We should spend exactly as much in each field as makes us first in that field.
\end{quotation}
If a competitor were to
establish a presence on the Moon, the US might worry that it was missing
out on something of which it was not aware. Here, I am reminded of
Seward's Folly, the purchase of Alaska, which in fact was both a strategic 
windfall and an economic bargain.

Over the many years required to return to the Moon, there are large risks
that could jeopardize the political will to continue the initiative. The
largest risk, in my estimation, is the US federal budget. The aging of the
``baby boom'' demographic of the USA and western Europe will soon greatly
increase the rate of retirements. How soon?  Beginning approximately
in 2010, which is 1945 (the end of World War II) plus 65 years (the
nominal age of retirement), wage earners that had been paying income
taxes and capitalizing equity markets will retire and begin to {\it
withdraw funds}. Political recognition of this impending financial bust
is evident; for example, on Jan 18, 2007, US Federal Reserve Chairman
Ben Bernanke testified on this specific topic before the Senate Budget Committee,
``We are experiencing what seems likely to be the calm before the storm....''

If somehow that bust does not materialize, or the lunar exploration
initiative somehow is immune to its effects, I expect a very substantial,
albeit largely intangible, benefit of lunar exploration may be to
bring the peoples of the world closer together, to save ourselves from
ourselves.  An excellent treatise on the latter topic was written at the
time of the Apollo program, ``The Tragedy of the Commons'' by Garrett
Hardin (1968). Hardin argues well that some problems do not have a
{\it technical} solution. He also points out that scientists, and often
policy makers as well, often assume that a technical solution exists and
fruitlessly seek one in their faith that one will eventually be found.
(Consider global climate change.)  Perhaps colonization of the Moon will
be a metaphor for solving Earth's geopolitical problems. For example, a
quasi-sustainable presence on the Moon wouldn't use fossil fuels; it would
use solar or nuclear power with a considerable emphasis on conservation.

Due to the large expense, in energy or dollars, of moving mass from
the surface of Earth to the surface of the Moon, 
nuclear power, in any of the various forms of radioisotope
thermoelectric generators, reactors, or explosives, may be convenient
for lunar exploration for electrical power, transmuting elements {\it in
situ}, or excavations.  However, any such convenience, i.e. of anything
brought from Earth, should be considered a negative compared to the
long-term strategic benefit of ``living off the land (regolith).'' For example,
solar power is readily and abundantly available on the lunar surface,
and a large, slow flywheel utilizing compacted regolith for mass
could provide for the variable power demands of human habitation and/or
store power through the lunar night away from the poles. An alternative
approach would be to bring from Earth a high-speed, precision flywheel
of relatively small mass or a chemical battery, but those are antithetical to
the strategic
benefit of utilizing lunar resources.  From the opposite perspective,
utilizing any water ice mined from lunar craters, for human consumption
and/or rocket fuel, could be myopic exploitation and destruction of an
important scientific resource. 

The benefits of knowledge always have had concomitant negative consequences:
\begin{quotation}
The idea that curiosity leads to disaster has an ancient pedigree. Pandora opened
the gods' box and let loose all the evils of the world; the descendants of Noah built
the Tower of Babel to reach heaven, but God scattered them and confounded their language;
Icarus flew so close to the sun that his homemade wings melted, and he fell into
the sea and drowned; Eve ate the apple of knowledge and was exiled from the Garden
of Eden; Faust traded his soul for sorcery and spent eternity in hell.... We believe
that curiosity is the beginning of knowledge and especially of science, but we know
that the application of science has led to disaster.  -- Finkbeiner (2006). 
\end{quotation}
Neils Bohr, as
attributed by Rhodes (1986), believed the knowledge of nuclear weapons
would ``foreclose the nation-state,'' by which he meant that knowledge 
of nuclear weapons was so powerful that its proliferation
would make the nation-state obsolete. Today, one might imagine any number
of Pandora's boxes equally well in that role, but none yet that have
been opened (at least not for long) outside governmental control.
It seems to me an inevitability akin
to the second law of thermodynamics that diffusion of knowledge will
occur, whether or not that diffusion is good for humankind. Technology
increasingly amplifies the power of an individual, or a small group of
individuals, for good or for evil.  

As Archimedes said of the lever,
\begin{quotation}
Give me a place to stand on, and I will move the Earth.
\end{quotation}
Here I suggest the lesser task of moving an asteroid, as an example of something
that seems entirely fanciful now, but which may not be so in decades hence.
Consider again the example given in the introduction, of
Captain Cook's voyage: at that time, it required government sponsorship, whereas
today's transportation
and technical infrastructure make replicating the feat entirely within the capacity
of a private individual.

Astronomers have suggested that they may give early warning of an asteroid
on a collision course with Earth.  Technologies useful to measuring
the orbit of such an asteroid with precision sufficient to enable the
prediction of an impending collision, and especially those technologies
useful to perturbing it to prevent a collision, are vaguely understood
and only potentially available to powerful nations today. Mostly, those nations
lack a specific motivation to act. However, in
some number of decades, what would have required the concerted effort of a nation
or nations might be accomplished by a group of individuals, but instead of
preventing a collision, that group could attempt clandestinely to create
a collision from what naturally would have been a near miss. Even a
credible threat to do so would be influential.  Presumably it is more
difficult to turn a near miss into a collision than vice versa, and
today we may take solace in our confidence that such a concept is entirely
impractical. 
My purpose is to illustrate the duality of
technology's amplification of power to individuals with a novel example potentially
relevant to the lunar exploration initiative and related technologies.
For those that reasonably consider the proposition patently absurd, I 
recite with intentional irony Margaret Mead,
\begin{quotation}
Never doubt that a small group of thoughtful, committed citizens can change the world.
Indeed, it is the only thing that ever has.
\end{quotation}

\section{Conclusions}\label{sec:concl}

An indirect method of imaging the oceans and continental boundaries of extrasolar planets
is outlined. Results from simulations of Earth-like extrasolar planets (McCullough 2007)
are presented in Section \ref{sec:glints} and demonstrate that the difference of fluxes in two orthogonal linear
polarizations is modulated by the planet's rotation, as it
alternately places continents or water at the location
on the sphere at which light from the star can be reflected specularly
to Earth. Section \ref{sec:context} digresses into the lunar exploration
initiative's potential impacts on society and vice versa.

\end{document}